# Performance Analysis of Sensing-Based Semi-Persistent Scheduling in C-V2X Networks

Amr Nabil, Vuk Marojevic, Komalbir Kaur, Carl Dietrich
Wireless@Virginia Tech, Bradley Dept. Electrical and Computer Engineering
Virginia Tech, Blacksburg, VA
{anabil, maroje, komalbir, cdietric}@vt.edu

*Abstract*—The 3[rd] Generation Partnership Project released the cellular vehicular-to-everything (C-V2X) specifications as part of the LTE framework in Release 14. C-V2X is the alternative to dedicated short range communications and both are specifically designed for V2X control signaling. C-V2X extends the device-to-device specifications by adding two more modes of operation targeting the vehicular environment in coverage and out of coverage of LTE base stations. Vehicle-to-vehicle communications (V2V) is established with Mode 4, where the devices schedule their transmissions in a distributed way employing sensing-based semi-persistent scheduling (SPS). Research is needed to assess the performance of SPS, especially in congested radio environments. This paper presents the first open-source C-V2X simulator that enables such research. The simulator is implemented in ns-3. We analyze the effect of the Mode 4 resource pool configuration and some of the key SPS parameters on the scheduling performance and find that the resource reservation interval significantly influences packet data rate performance, whereas resource reselection probability has little effect in dense vehicular highway scenarios. Our results show that proper configuration of scheduling parameters can significantly improve performance. We conclude that research on congestion control mechanisms is needed to further enhance the SPS performance for many practical use cases.

*Keywords*—C-V2X, LTE, Mode 4, Open-Source Simulator, Performance Analysis, Semi-Persistent Scheduling.

I. INTRODUCTION

The long-term evolution (LTE) standard was originally introduced for providing mobile broadband service to cellular subscribers. LTE has considerably extended its scope since its first release, Release 8. The Third Generation Partnership Project's (3GPP's) Release 12 introduced Proximity Services (ProSe) for Device-to-Device (D2D) Communications. D2D enables the quick exchange of data over short distances via a direct link between nodes. This offers an efficient way to bypass the LTE base station or eNB and thus offload the eNB traffic. Besides content sharing, a D2D user equipment (UE) can act as a relay for another device with a poor connection to the eNB and, therefore, D2D can be used to extend of cellular network coverage. Two modes, Modes 1 and 2, have been defined for centralized and distributed scheduling of UE transmissions. Centralized scheduling occurs at the eNB (in-coverage mode), whereas distributed scheduling is carried out by the D2D UEs themselves, with no need to be in the coverage area of an eNB (out-of-coverage mode). The operational principle of Modes 1 and 2 is battery life improvement of mobile devices.

Vehicular communications has other constraints that cannot be accommodated with ProSe. Specifically, the high latencies of ProSe are not suitable for vehicular communications, where packet delays or packet losses can have severe and life-threatening consequences. 3GPP Release 14 therefore extends the ProSe functionality by adding two new modes, Modes 3 and 4, for cellular-vehicle-to-everything (C-V2X) connectivity. C-V2X encompasses vehicle-to-vehicle (V2V), vehicle-to-pedestrian (V2P), vehicle-to-infrastructure (V2I) and vehicle-to-network (V2N). V2V establishes the direct link between vehicles that are in close proximity to one another. Basic Safety Messages (BSMs) and event-triggered messages are transmitted for collision avoidance. V2V also enables platooning and cooperative automated driving. V2P establishes the communications protocol between vehicles and pedestrians for pedestrian safety. V2I implies the communications with roadside units and allows making information about local road and traffic conditions readily available to vehicles. V2N enables commercial services by providing access to data stored in the Cloud.

Modes 3 and 4 have been designed to satisfy the latency requirements and accommodate high Doppler spreads and high density of vehicles for C-V2X communications. The maximum allowed latency varies between 20 and 100 ms, depending on the application. Mode 3 uses the centralized eNB scheduler. The vehicular UE and eNB use the Uu interface to communicate. Mode 4 uses the new PC5 interface, which offers direct LTE sidelink (SL) between two vehicular UEs. It employs distributed UE scheduling.

The emphasis of this paper is on C-V2X Mode 4, which offers an alternative to dedicated short-range communications (DSRC). DSRC has been standardized for many years. It is based on IEEE 802.11, adapted to vehicular environments. Several simulations and trials have been completed and reported in open literature [3]. C-V2X is rather new; however, a few research articles have recently appeared. Most prominently, Molina-Masegosa et al. [4][5] provide an introduction to C-V2X Mode 4 and present results that compare random resource selection against distributed scheduling and the effect of retransmission in an urban scenario. They also compare C-V2X against DSRC.

This paper analyzes the semi-persistent scheduling (SPS) of C-V2X, which is used by vehicular UEs operating in Mode 4 to self-allocate resources for BSM transmission. We provide a tutorial-like overview of its operation and analyze the effect of the different parameters and parameter choices. To that end, we

This work was in part supported by Ford Motor Company through the University Alliance Program.

developed a 3GPP compliant C-V2X Mode 4 simulator in ns-3, extending the LTE D2D simulator [6] developed by the National Institute for Standards and Technology (NIST).

Section II describes C-V2X Mode 4 and the SPS operation. Section III introduces the first open-source C-V2X simulator and the parameters that we considered for the numerical analysis. We analyze the effect of the resource pool configuration and some of the key SPS parameters on the scheduling performance and show that proper configuration of parameters can significantly improve performance (Section IV). We conclude the paper by providing an outlook on future research directions on scheduling and congestion control that are much needed and enabled by the simulator (Section V).

## II. C-V2X MODE 4

C-V2X Mode 4 operates without infrastructure support, although the UEs could be in eNB coverage. Resources are shared with the LTE uplink. Both LTE duplexing modes, time and frequency division duplexing, are supported. Mode 4 uses a specific resource pool configuration and SPS to select and reserve resources for transmission.

### A. Sidelink Control Information

The SL Control Information (SCI) Format 1 was introduced in 3GPP Release 14 (D2D uses SCI Format 0). It is transmitted over the Physical SL Control Channel (PSCCH), which carries the information related to the transmission of data over the Physical SL Shared Channel (PSSCH). The SCI Format 1 informs the receiving vehicular UEs about the resource reservation interval, frequency location of initial transmission and retransmission (up to one retransmission is supported in Mode 4), time gap between initial transmission and retransmission ($SF_{gap}$), and modulation and coding scheme (MCS) used to modulate the data transmitted over the PSSCH. The PSSCH carries the BSMs and event-triggered messages.

### B. Physical Layer

C-V2X employs singe-carrier frequency division multiple access (SC-FDMA). The control information is QPSK modulated, whereas QPSK or 16-QAM can be used to modulate the data. Only normal cyclic prefix is allowed for Mode 4 and the last symbol in a subframe is used as a guard period.

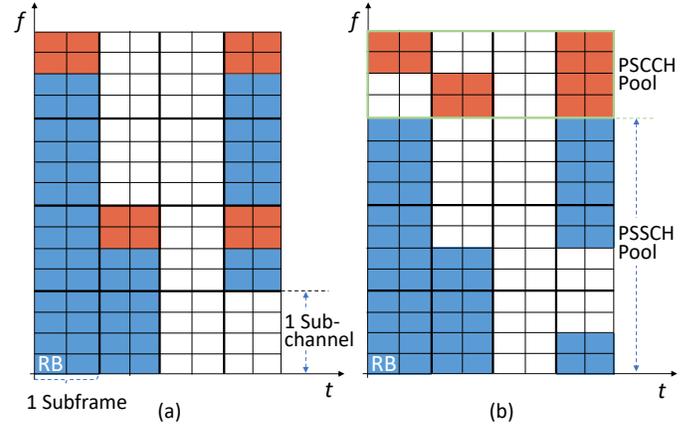

**Fig. 1.** Adjacent (a) and nonadjacent (b) PSCCH and PSSCH.

When compared to Modes 1 and 2, the number of Demodulation Reference Signal (DMRS) subframes has been increased from two to three per resource block (RB) for the Physical SL Broadcast Channel (PSBCH) and to four for the PSCCH and PSSCH in order to account for the high Doppler spread at absolute vehicle speeds of up to 250 km/h and relative speeds of up to 500 km/h [1].

### C. Resource Pool

C-V2X allows channel bandwidths of 10 and 20 MHz. A channel is further divided into subchannels in the frequency domain and subframes in the time domain. Each subchannel is further divided into RBs. Each RB is defined as 12 subcarriers by 1 slot, or 180 kHz by 0.5 ms, as in other LTE operational mode. The number of RBs per subchannel can vary. The time granularity for message scheduling is the subframe or transmission time interval (TTI) of 1 ms.

Two configurations can be employed for the PSCCH and PSSCH transmission. In the adjacent configuration (Fig. 1a), the SCI and data are transmitted in adjacent RBs, occupying one or several subchannels. The SCI is transmitted in the first two RBs in the reserved subframe followed by the data. In the nonadjacent configuration, two separate pools are configured for the PSCCH and PSSCH (Fig. 1b).

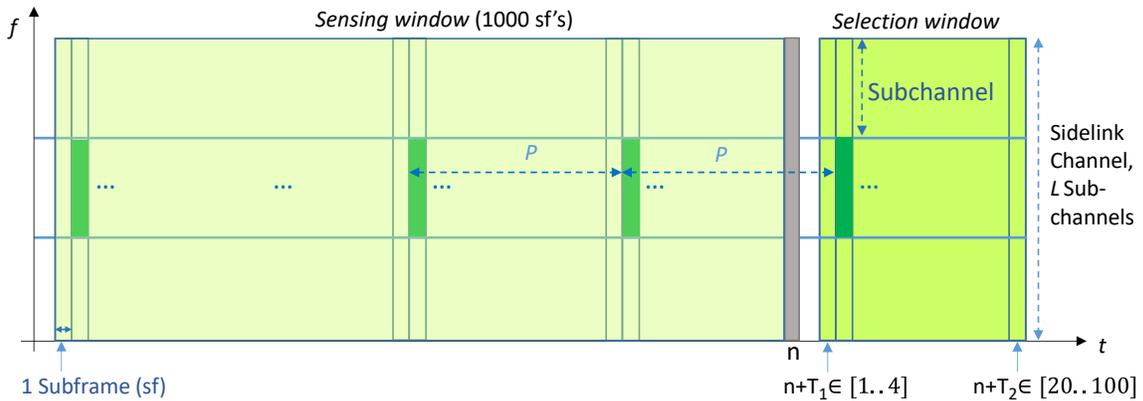

**Fig. 2.** Sensing and selection windows for semi-persistent scheduling in C-V2X.

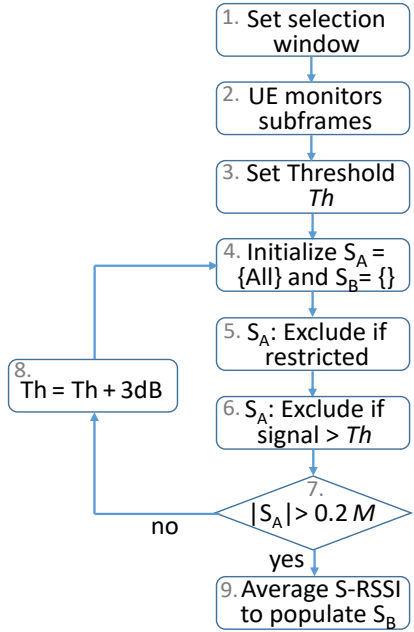

**Fig. 3.** Semi-persistent scheduling procedure in C-V2X.

*D. MAC: Semi-Persistent Scheduling*

SPS was introduced in LTE for supporting services that require deterministic latency, such as voice. Mode 4 adopts this concept and uses sensing to determine suitable semi-persistent transmission opportunities, i.e. the set of subframes and subchannels for transmission. A candidate single-subframe resource consists of from one up to $L$ contiguous subchannels in a single subframe, depending on the message size. The UE shall select a set of candidate single-subframe resources within the selection window that spans a number of subframes (Fig. 2) and contains $M$ single-subframe resources. Parameters $T_1$ and $T_2$ define the selection window. $T_2$ is determined as a function of the latency requirement [9].

Fig. 3 illustrates the SPS procedure, which is described in continuation and adapted from [1]:

- The UE continuously monitors subframes and takes notes of decoded SCI and SL received signal strength indicator (S-RSSI) measurements (2.). It considers the last 1000 subframes for selecting candidate single-subframe resources as follows.
- The UE sets the signal power threshold $Th$ and initializes sets $S_A$ and $S_B$ as complete and empty sets (3., 4.).
- The UE excludes all candidate single-subframe resources if the UE has not monitored the corresponding resource in the sensing window (5.).
- Out of all remaining candidate resources, those that are reserved or where a PSSCH reference signal received power (RSRP) measurement is higher than $Th$ are excluded (6.).
- If the remaining single-subframe candidate resources are less than 20% of the total resources, the threshold is increased by 3 dB (7., 8.) and the two-step exclusion process is repeated.
- Otherwise, the linear average of S-RSSI is computed and those 20% of single-subframe candidate resources with the smallest average S-RSSI in set $S_A$ are moved to set $S_B$ (9.), which is reported to higher layers.
- The MAC then randomly selects a candidate resource of $S_B$ for the first transmission.
- If the number of hybrid automatic repeat request (HARQ) transmissions is two, then for the set of selected transmission opportunities $t_{n+k}$, where $k = 0, 1,.., K$, another set of transmission opportunities $t_{n+k+j}$ is also selected where $j \neq 0$ and $-15 \leq j \leq 15$.

The SPS grant persists for the next *ResourceCounter* (RC) packet transmissions, where RC is randomly selected in [2]:

- [5, 15] if resource reservation interval $\geq$ 100 ms,
- [10, 30] if resource reservation interval = 50 ms, or
- [25, 75] if resource reservation interval = 20 ms.

After *RC* reaches 0, the probability *P* to keep the previous resources is between 0 and 0.8, depending on the configuration.

### III. OPEN-SOURCE SIMULATOR AND SIMULATION SETUP

*A. Simulator*

We implemented C-V2X Mode 4 procedures including SPS in the Network Simulator 3 (ns-3). The simulator was originally modified by NIST [6], [7] to support Modes 1 and 2 of the LTE SL for D2D. The NIST team integrated their implementation of SL Modes 1 and 2 with ns-3. We made the required modifications to NIST data structures and functions, and reused the provided interfaces with other LTE modules. Although we reused and modified existing code where possible, we developed original code to implement the core functions and data structures required for the correct operation of SL transmission under Mode 4.

The sensing-based SPS algorithm is mandated in Release 14 of the LTE specification as a unique feature of Mode 4. Fig. 4 shows the block diagram of the MAC and PHY layers of the modified ns-3 LTE D2D simulator. At the PHY layer (LteUePhy), we added a sensing module that measures the signal strengths of all received packet on the corresponding resource blocks and stores this data in the sensing data structure. This data structure keeps the measured signal strength of the latest subframes, according to the sensing window. For each new subframe, the new data is stored and the oldest sensing data discarded. The SPS module is implemented in the MAC layer (LteUeMac). It uses the sensing data structure as an input to determine the resource allocation as described in Section II.

*B. Simulation Setup*

*1) Vehicular scenario:* We consider a freeway scenario where each vehicle within a set of vehicles is moving with the same average speed. In our simulations, we consider two values for the speed—70 and 140 km/h—as suggested in [8]. The freeway consists of six lanes, three in each direction. We consider a two kilometer road segment. The initial location of vehicles on the road follows a spatial Poisson process. The density of vehicles in each lane is obtained by taking the inverse of 2.5 s mul-tiplied by the average vehicle speed. For example, at 70 and 140 km/h average speed, the average vehicle densities are 20.6 and 10.3 vehicles per km, respectively. A warp around is applied at the end of the 2 km highway segment. This means that when a vehicle reaches one end of the road segment, it is moved immediately to the other end of the road segment and is kept in the same lane.

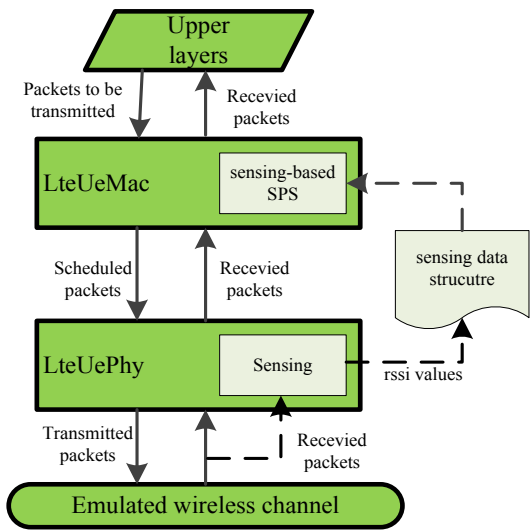

**Fig. 4.** Block diagram of the LTE C-V2X ns-3 simulator.

*2) RF channel model:* The path loss is determined using the line of sight model in WINNER+ B1. The antenna height at each vehicle is set to 1.5 m. The path loss for distances less than 3 m is set using the value at a transmitter-receiver separation of 3 m. The shadowing distribution is Log-normal with a standard deviation of 3 dB. Shadowing spatial correlation is implemented with a decorrelation distance of 25 m, as suggested in [8]. Each vehicle is transmitting at 23 dBm in the 5.9 GHz band. We assume perfect synchronization among all vehicles.

*3) Message and frame configuration:* Each vehicle broadcasts BSMs with a pattern of four short messages, followed by one long message, i.e. {190, 190, 190, 190, 300} bytes [8]. The time interval between two subsequent packets is initially set to 100 ms. The network is configured to use adjacent PSCCH + PSSCH subchannelization scheme, as illustrated in Fig. 1a. Each subframe extends in frequency across a finite number of $L$ subchannels, where each subchannel consists of 12 RBs or 2.16 MHz. The packets are ½ rate encoded and modulated using QPSK. Each short packet fits in one subchannel, whereas a long packet occupies two subchannels. Parameters $T_1$ and $T_2$ of Fig. 2 are set to 4 and 20, respectively. The resource reservation interval, probability of resource reselection, and the number of available subchannels for SL transmission are the variables for our simulations. In continuation we study the effect of each of these parameters on the overall network performance.

## IV. NUMERICAL RESULTS

In this section, we analyze the performance of the sensing-based SPS for C-V2X networks. We study the effect of (*A*) the number of subchannels allocated to SL communications, (*B*) the resource reservation interval, and (*C*) the probability of reselecting the resources on the packet delivery ratio (PDR). The PDR is defined as the ratio between the number of successfully decoded packets to the total number of transmitted packets from all transmitters. A packet is considered to be received successfully if there is no overlap between the subchannel(s) over which the packet has been transmitted and the subchannel(s) that carry any other transmitted packet in the same subframe. The packet count is based on each Tx-Rx pair. For example, if a UE broadcasts the same message to ten other UEs, the total number of packets is ten.

### A. Effect of the number of dedicated subchannels

In the first set of experiments, we study the effect of the number of available SL subchannels on the PDR for two different average UE speeds, which translate to different vehicle densities. As shown in Fig. 5, the PDR value when the UE speed is 70 km/h is always lower than when the average speed is 140 km/h. This is so because of the 2x higher vehicle density at 70 km/h and, as a result, a larger number of collisions is experienced. Recall that an overlap between the subchannels reserved by two or more transmitting UEs causes the packets to collide at the intended receiver.

As expected, the larger number of SL subchannels that is available to the UEs to select the resources from, the lower the probability that more than one UE chooses the same resource. Consequently, the number of successfully received packets, and hence the PDR, will be higher.

### B. Effect of the SPS resource reservation interval

Here we discuss the effect of the resource reservation interval on the PDR. Since the assumed inter-arrival time of application packets is 100 ms, we consider resource reservation intervals of 100, 200, …, 1000 ms. The corresponding values of X in the specifications and in Fig. 6 are 1, 2,…, 10. As shown in Fig. 6, the PDR increases as X increases. This can be explained as follows. As the resource reservation interval increases, the number of used resources by a UE within a given period decreases. Consequently, the probability of reserving the same resource by more than one UE decreases and the overall PDR increases. It should be noted that increasing the resource reservation degrades the data rate and increases the packet delay. Hence, there is a tradeoff between these performance measures.

### C. Effect of probability of resource reselection

In the next set of experiments we analyze the PDR as a function of the probability of going through the SPS process for selecting new resources when the RC reaches 0. This resource reselection probability is defined as 1–$P$ by 3GPP [2]. The number of available subchannels for SL transmission is set to four. We use two values for the resource reservation interval (100 and 500 ms). Fig. 7 shows the results. We do not notice any significant variation in the PDR when changing $P$ between 0 and 0.8. This observation ca be explained as follows. Initially, UE1 chooses its resources using the sensing-based SPS. On the other hand, other UEs will avoid choosing the same resources which have been reserved by UE1. The number of UEs in the network is relatively large. This will result in reserving most of the resources available in a selection window. Then, UE1 will find a limited amount of available resources while re-executing the SPS algorithm when its RC expires (which happens with a probability 1–$P$). Consequently, it is probable that UE1 will end up choosing the same resources again and this would be equivalent to maintaining the resource reservation without executing the SPS algorithm (which happens with a probability $P$).

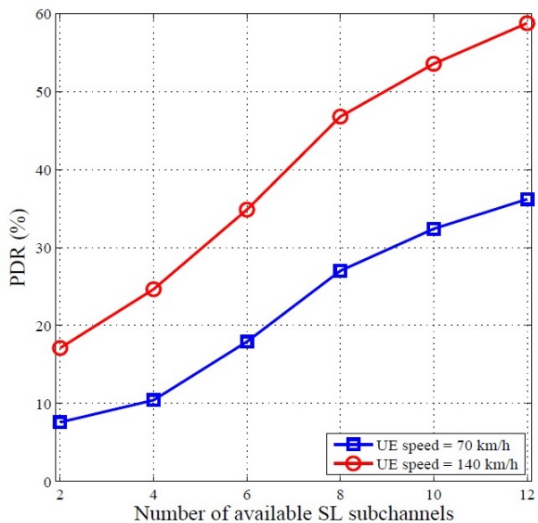

**Fig. 5.** Effect of the number of SL subchannels.

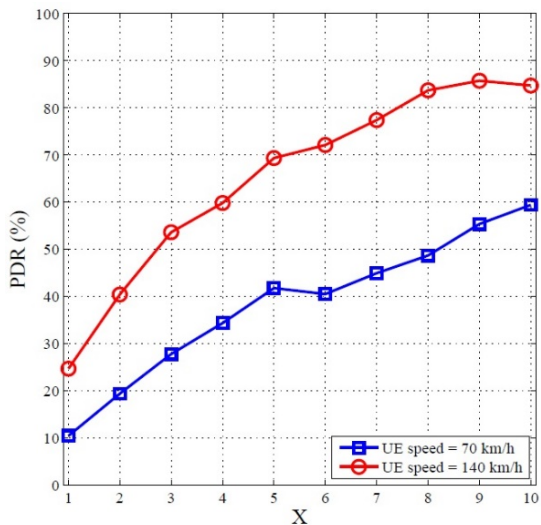

**Fig. 6.** Effect of the resource reservation interval of 100X ms.

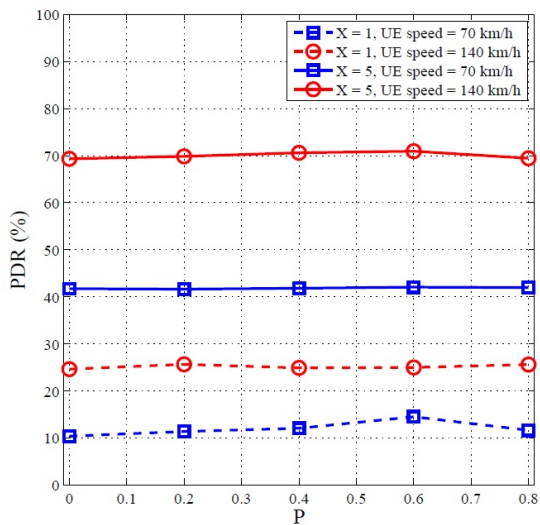

**Fig. 7.** Effect of the resource reselection probability 1–$P$.

## V. CONCLUSIONS

This paper provided an overview of LTE C-V2X Mode 4 with emphasis on SPS to enable vehicle-to-vehicle communications in the absence of cellular infrastructure. We presented our ns-3 simulator implementation and analyzed the performance of SPS in two data traffic scenarios and three SPS configuration parameters. Our results show that the PDR improves as the number of available subchannels for SL communications increases. Moreover, the resource reservation interval significantly influences the achievable PDR in dense networks studied here. The PDR increased from 10-25% to 60-85% when we increased the resource reservation interval from 100 to 1000 ms. Increasing this interval, however, reduces the data rate and increases the packet delay. A compromise is thus needed to select the optimal parameter value. Finally, our simulations revealed that because of the high density of the network, the probability of reselecting the resources using SPS procedure has almost no effect on the overall performance.

In future work we will study different scenarios (other than the freeway) to further evaluate the SPS performance. Our results have shown that the performance of SPS in highly-congested networks is not satisfactory and has room for improvement. Therefore, we conclude that research on suitable congestion control mechanisms is needed to enhance the SPS performance in a variety of practical C-V2X scenarios, many of which are likely to be congested.